\documentclass[10pt,a4paper,twocolumn,notitlepage]{article}
\usepackage{amsmath,amssymb,amsfonts}

\usepackage{graphicx}
\usepackage[margin=2cm]{geometry}
\usepackage{caption,subcaption}
\usepackage{tikz}
\usepackage{textcomp}
\usepackage{babel}
\usepackage{authblk}
\usepackage{enumitem}
\usepackage{amssymb}
\usepackage{xcolor}

\PassOptionsToPackage{normalem}{ulem}
\usepackage{ulem}
\providecolor{added}{rgb}{1.,.0,0.}
\newcommand{\todo}[1]{{}}

\begin{document}

\title{Accelerometry-based classification of circulatory states during out-of-hospital cardiac arrest}
\author[1,2]{Wolfgang J. Kern}
\author[2,3,4]{Simon Orlob}
\author[5,6]{Andreas Bohn}
\author[3]{Wolfgang Toller}
\author[4,7,8]{Jan Wnent}
\author[4,7]{Jan-Thorsten Gräsner}
\author[1,2]{Martin Holler}

\affil[1]{Institute of Mathematics and Scientific Computing, University of Graz, Heinrichstraße 36, 8010 Graz, Austria}
\affil[2]{BioTechMed-Graz, Graz, Austria}
\affil[3]{Medical University of Graz, Department of Anesthesiology and Intensive Care Medicine, Division of Anesthesiology for Cardiovascular and Thoracic Surgery and Intensive Care Medicine, Graz, Austria}
\affil[4]{University Hospital Schleswig-Holstein, Institute for Emergency Medicine, Kiel, Germany}
\affil[5]{City of Münster Fire Department, Münster, Germany}
\affil[6]{University Hospital Münster, Department of Anesthesiology, Intensive Care and Pain Medicine, Münster, Germany}
\affil[7]{University Hospital Schleswig-Holstein, Department of Anaesthesiology and Intensive Care Medicine, Kiel, Germany}
\affil[8]{University of Namibia, School of Medicine, Windhoek, Namibia}

\onecolumn
\maketitle 

\begin{abstract}
Objective: Exploit accelerometry data for an automatic, reliable, and prompt detection of spontaneous circulation during cardiac arrest, as this is both vital for patient survival and practically challenging. Methods: We developed a machine learning algorithm to automatically predict the circulatory state during cardiopulmonary resuscitation from 4-second-long snippets of accelerometry and electrocardiogram (ECG) data from pauses of chest compressions of real-world defibrillator records. The algorithm was trained based on 422 cases from the German Resuscitation Registry, for which ground truth labels were created by a manual annotation of physicians. It uses a kernelized Support Vector Machine classifier based on 49 features, which partially reflect the correlation between accelerometry and electrocardiogram data. Results: Evaluating 50 different test-training data splits, the proposed algorithm exhibits a balanced accuracy of 81.2\%, a sensitivity of 80.6\%, and a specificity of 81.8\%, whereas using only ECG leads to a balanced accuracy of 76.5\%, a sensitivity of 80.2\%, and a specificity of 72.8\%. Conclusion: The first method employing accelerometry for pulse/no-pulse decision yields a significant increase in performance compared to single ECG-signal usage.
Significance: This shows that accelerometry provides relevant information for pulse/no-pulse decisions. In application, such an algorithm may be used to simplify retrospective annotation for quality management and, moreover, to support clinicians to assess circulatory state during cardiac arrest treatment.
\end{abstract}
\let\thefootnote\relax\footnotetext{"(c) 2023 IEEE. Personal use of this material is permitted. Permission from IEEE must be obtained for all other uses, in any current or future media, including reprinting/republishing this material for advertising or promotional purposes, creating new collective works, for resale or redistribution to servers or lists, or reuse of any copyrighted component of this work in other works."}
\let\thefootnote\relax\footnotetext{"This article is accepted for publication in \textit{IEEE - Transactions on Biomedical Engineering} and can be found under doi: 10.1109/TBME.2023.3242717.https://ieeexplore.ieee.org/document/10038473 "}
\let\thefootnote\relax\footnotetext{https://ieeexplore.ieee.org/document/10038473}

\twocolumn
\section{Introduction}
\label{sec:introduction}

Cardiac arrest is one of the leading causes of death in the western world  \cite{Grasner:2021} with more than 400 000 resuscitation attempts by emergency medical services (EMS) per year in Europe alone \cite{Grasner:2021}. High-quality cardiopulmonary resuscitation (CPR) - ensuring minimal circulation - along with defibrillation is the basic treatment EMS provide for patients with cardiac arrest.
The objective of the treatment is a restoration of spontaneous circulation. However, recognizing a return of spontaneous circulation (ROSC) during brief rhythm checks is still a demanding task. Currently, health-care providers check for mechanical activity of the heart by manual palpation of central pulses when the electrocardiogram (ECG) shows a potential perfusing rhythm. Manual pulse palpation has two big disadvantages: It often leads to long interruptions of CPR \cite{ochoa:1998} and is highly error-prone \cite{eberle:1996}. Erroneous identification of spontaneous circulation leads to inadequate treatment by delaying chest compression and, consequently, decreases the patient's survival probability \cite{christenson:2009} due to prolonged no-flow time. Physicians additionally employ endtidal $\text{CO}_2$-concentration levels and their trend as further information to augment the assessment of the circulatory state \cite{Soar:2021}. 
Hence, in the last two decades various algorithms for organized ECG rhythms were proposed to differentiate pulse generating perfusing rhythms (PR) from pulseless electric activity (PEA). These algorithms employ ECG \cite{Elola:2019,Elola:2019a,Kwok:2020}, thoracic impedance (TI) \cite{Cromie:2008,Elola:2019b}, combinations of those \cite{Risdal:2008,Alonso:2016,Ruiz:2018}, and additionally capnography (CO2) \cite{Elola:2019c,Elola:2020} or photoplethysmography \cite{Wijshoff:2015} for classification, or are based on Doppler ultrasound \cite{cohen:2022}. 
Some works \cite{Elola:2020} subclassified PEA into pseudo-PEA, and true-PEA, which differ in terms of cardiac output which is insufficient for pseudo-PEA and lacking for true-PEA, respectively. Even though, the performance of this classifiers increased in recent years, the problem of automatic PR/PEA classification is far from being solved. \\
In cardiology, data from accelerometers placed on the patient’s chest, called seismocardiography, are investigated for many years (see e.g. \cite{Taebi:2019} and references therein) on healthy persons as well as on patients with cardiovascular diseases allowing non-invasive and continuous measurements. Due to the more controllable setting, seismocardiography is used to determine not only the circulatory state, but detailed heart rates and cardiac time intervals. We use the term "seismocardiography" to describe these detailed measurements throughout this paper. 

Although accelerometers are placed on the patient's chest as feedback devices for the EMS delivering chest compressions during CPR, these accelerometry data were only recently investigated on their capability to provide complimentary information about the patient's circulatory state. A proof of concept of this idea was provided in porcine models \cite{Wei:2017} and with smartphone-accelerometry in humans \cite{Lee:2021}, where the latter one focused only on the discrimination of the signals by blinded observers, whereas \cite{Wei:2017} proposed a classifier based only on a single feature. Since we are interested only in pure detection of any mechanical cardiac movement without further investigation, we use the term "accelerometry" (ACC) for this process.  \\
This study proposes a comprehensive machine-learning algorithm employing both real-world ECG and ACC signals to predict the circulatory state during cardiac arrest, thereby proving the usefulness of accelerometers for circulatory state classification and potentially supporting clinicians in decision making in the field.

\section{Methods}
\subsection{Data collection and labeling}
\label{ssec:Data}
This study was approved by the ethics committee of the University of Kiel (Ref. no.: D 421/21) and the scientific advisory board of the German Resuscitation Registry (Ref. no.: AZ 2021-03). A sample of $871$ cases with defibrillator records, all ZOLL X-Series (ZOLL Medical Corporation, Chelmsford, Massachusetts, United States), from the years $2013$ to $2021$ was obtained from the German Resuscitation Registry. All recordings were annotated retrospectively by an experienced physician (SO), utilizing an interactive, web-based plotting tool that we developed for a previous study using jupyter notebooks \cite{jupyter}. Cases recorded in the year $2020$ were annotated independently by two experienced physicians, and dissenting annotations were resolved in consensus.  The annotation consists of the determination of the start and the end of the resuscitation episode, including all ROSCs and rearrests which occurred in this interval \cite{kramerjohansen:2007}. The physicians used ECG, ACC and, if available, capnography and non-invasive blood pressure for the labeling. Ambiguous situations were annotated assuming the treating physician followed the guidelines faultlessly. Nevertheless, obtuse intervals could be excluded from the subsequent analysis by a respective label from the annotators. These annotations form the ground truth in our data set. \\
After resampling ECG and ACC to $f_s=250$ Hz for all cases, only the parts of the recordings with simultaneously given ACC, ECG, and capnography signals were used further. We require the presence of capnography data since these provide at least some hemodynamic information for the retrospective annotation process. Furthermore, the signal after the last recorded period of chest compressions was excluded since it contains either mainly the patient's transport to the hospital with potentially large artifacts or a low amplitude signal in ECG and ACC after death pronouncement, where circulation classification is clinically irrelevant. For the remaining signal parts, the algorithm described in \cite{Orlob:2021,Kern:2021} was used to determine the periods where no chest compressions are present. From these periods, we extracted snippets with a length of $4$ seconds, each one containing ACC and ECG signals and a label ('Cardiac Arrest' (AR) or 'Spontaneous Circulation' (SC)). The snippets are extracted in an overlapping way, cutting a $4$-s-snippet every $2$ seconds. These snippets form the database of our algorithm.
We used the trained algorithm of \cite{Rad:2017} to determine the rhythm of each snippet (either Asystole (ASY), Ventricular Fibrillation (VF), Ventricular Tachycardia (VT), PEA or PR). Since both PEA and PR exhibit an organized looking rhythm differing only in cardiac output, we merge these to ryththm classes to one Organized Rhythm (ORG) class for most of our analysis.

\subsection{Preprocessing}
\label{ssec:preprocessing}
We denote the ACC and ECG signals of a snippet with $N$ sample points as $a=(a_n)_n$ and $e=(e_n)_n$, where the indices $n$ are such that $-N/2 \leq n < N/2$ respectively. We use $\overline{d}$ to denote the mean $\overline{d}=(1/N) \sum_{n=-N/2}^{N/2-1} d_n$ of some signal $d$.\\
For all snippets we shift $a$ and $e$ so that they exhibit $\overline{a}=\overline{e}=0$. 
To improve data quality we aim to discard snippets containing e.g. shocks, baseline changes in the ACC signal and movement and transport artifacts by imposing the following constraints in a prefiltering step: Constraining $\max |a_n|<20$, $\max |e_n|<2.5~\text{mV}$ removes data with high amplitude noise, whereas  $\max(|a_n|) / \overline{|a_n|}<25$  and $\max(|e_n|) / \overline{|e_n|}<35$ allow to identify and omit snippets with a sharply peaked artifacts.

\subsection{Feature extraction}

We extract several features from the ECG-signal $e$ and acceleration signal $a$ to use them as an input for the machine learning classifier, which were partly taken from the literature and partly developed by ourselves. The latter ones are introduced subsequently. We use the root mean square $v_1=\sqrt{\overline{a_n^2}}$ and $v_2=\sqrt{\overline{e_n^2}}$ as features. To characterize the rhythmicity we employ the autocorrelation $z_d$ of a real-valued signal $d$
\begin{align}
z_d (k)= \frac{ \sum_{n=-N/2}^{N/2-1} d_n d_{n+k}}{ \sum_{n=-N/2}^{N/2-1} d_n^2 }, -N/2\leq k < N/2,
\end{align}
where we zero-pad $(d_n)_n$ outside the defined range $ -N/2 \leq n < N/2$.  As usual $z_d(0)=1$ holds, which we call the trivial maximum. We search for the largest nontrivial local maximum $z_d (\tilde{n}_d) ,\tilde{n}_d\neq 0$. Its value $z_d (\tilde{n}_d)$ describes the rhythmicity of the signal, whereas $\tilde{n}_d/f_s$ gives the time shift inducing highest self-similarity.  
We use the $v_3=z_a (\tilde{n}_a)$ and $v_4=z_e (\tilde{n}_e)$ as further input features. 
In order to describe the interdependence of $a$ and $e$ we use three different approaches:

\noindent \textbf{1)} We compute the Fourier Transform of $a$ and $e$ and take the absolute values of the coefficients in the frequency band between $1$ Hz and $20$ Hz: $(s_a)_{i=1}^m$ and $(s_e)_{i=1}^m$. Then we take the correlation of these two vectors as a feature
\begin{align}
v_5=\sum_{i=1}^m s_{a_i} s_{e_i} / \sqrt{\sum_{i=1}^m s_{a_i}^2 \sum_{i=1}^m s_{e_i}^2}.
\end{align}

\begin{figure}
\centering
\includegraphics[trim=3.5cm 2cm 3cm 2.5cm,clip,width=.5 \textwidth]{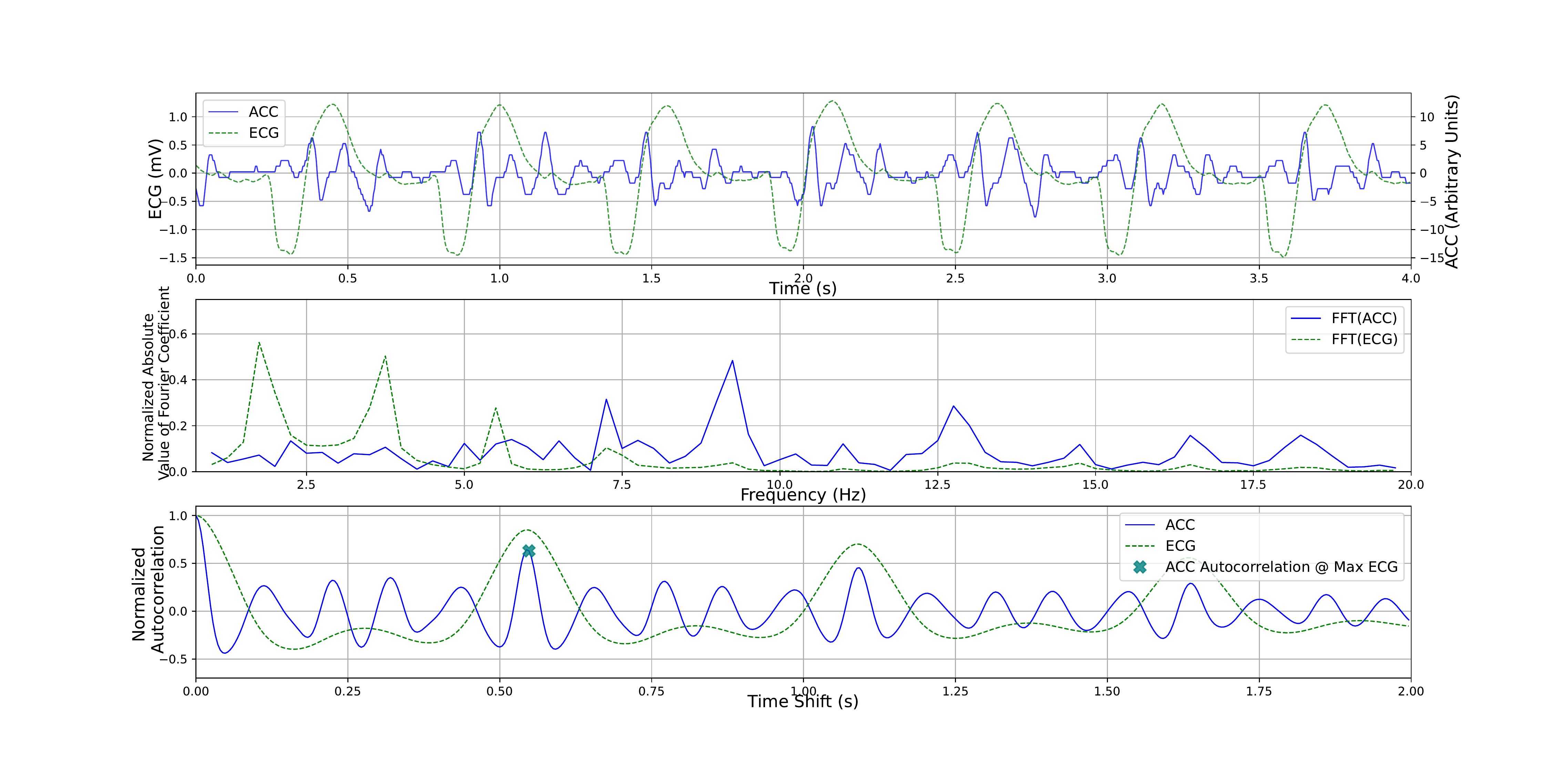}
\caption{Exemplary Snippet. ACC (blue) and ECG (green) signals are shown in the first subplot, their spectra in the second subplot. The autocorrelation is shown in the lowermost subplot. Even though one can see apparent acceleration excitation at the time of the QRS-complexes in ECG, the spectra exhibit different harmonics, leading to a small $v_5=0.0595$. In contrast, the features based on the autocorrelation yield $v_6=0.631$ and $v_7=-2920$ at the pronounced maximum at $0.55$ s time shift.}
\label{fig:SpOV_not_sufficient}
\end{figure}

\begin{figure}
\centering
\includegraphics[trim=3.5cm 1.5cm 3cm 2.5cm,clip,width=.5 \textwidth]{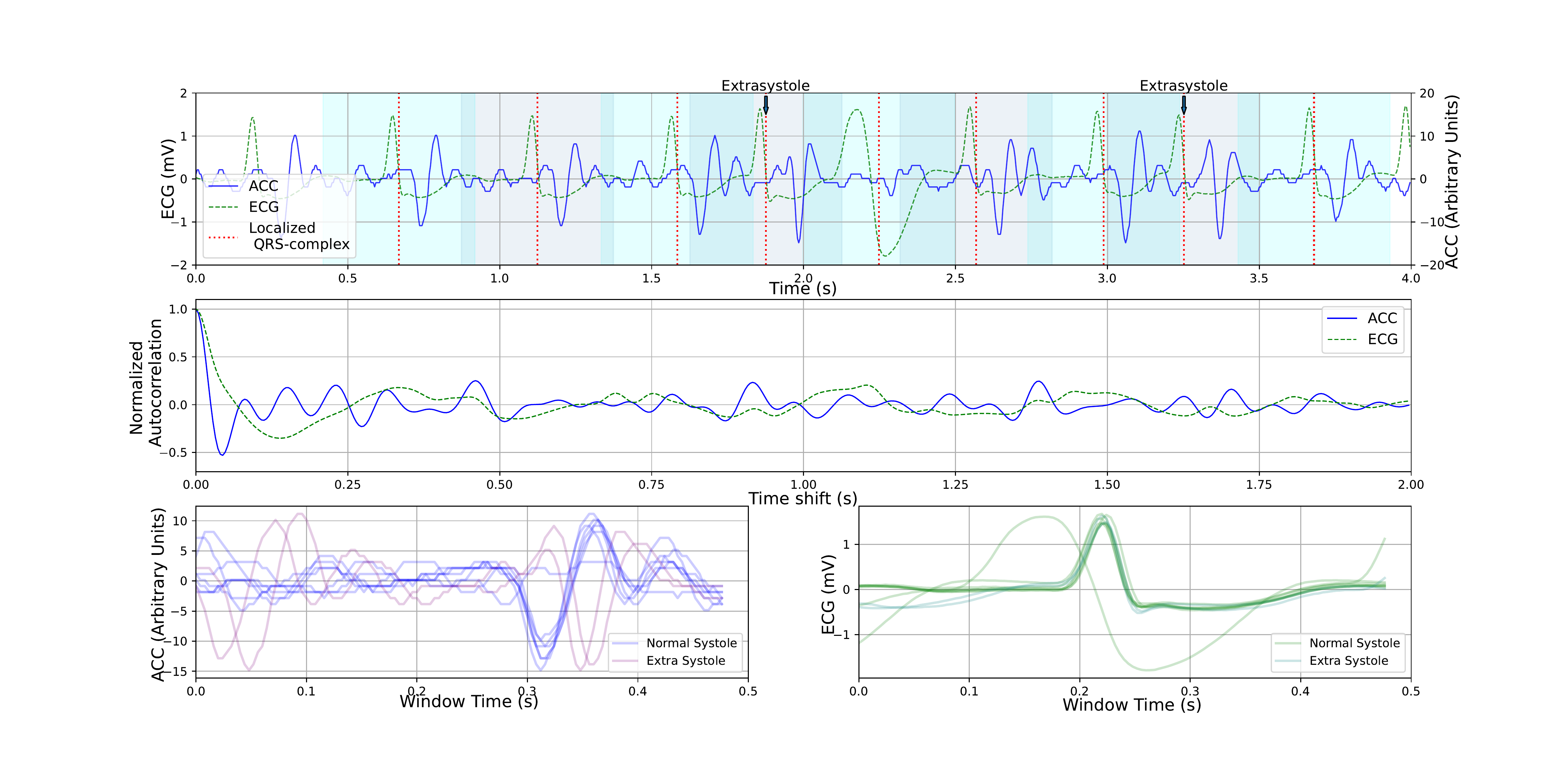}
\caption{Exemplary Snippet. ACC (blue) and ECG (green) signals are shown in the first subplot, their autocorrelation in the second subplot. Even though one can see clear ACC excitations at the time of the QRS-complexes in ECG, the autocorrelation maxima remain small due to arrhythmia: $v_6=0.004$ and $v_7=214$. Windows of fixed size localized around QRS-complexes are shown alternatingly in different pale shades of blue. The ACC and ECG signals during these windows are superimposed in the lowermost subplots, which reveal the high similarity. The features based on the correlation of the windowed signals yield $v_{8}=0.770$, $v_{9}=0.917$ and $v_{10}=0.840$. Furthermore, the two extrasystoles are marked with a black arrow in the uppermost subplot and their windowed signals are shown with a different color in the overlaps below. The different acceleration patterns are clearly visible.}
\label{fig:Arhy_Snippet}
\end{figure}

With this feature we aim to identify similar frequencies in both signals. However, as can be seen in Fig. \ref{fig:SpOV_not_sufficient}, there are situations where different harmonics contribute to ECG and ACC spectra, leading to small values of $v_5$ even in case of clearly visible acceleration excitations following the QRS-complexes. Thus, this feature is insufficient to describe the interdependence. \\
\noindent \textbf{2)} We employ the autocorrelation of the two signals (as shown in Fig. \ref{fig:SpOV_not_sufficient}) to characterize the dependency of both signals further. We determine the shift $\tilde{n}_e$ at which the nontrivial maximum of the ECG autocorrelation occurs and evaluate $v_6:=z_a(\tilde{n}_e)$. If the acceleration signal is influenced by the apex beat against the chest wall, we will observe simultaneous, rhythmic responses of the acceleration signal to this excitation represented by the QRS-complexes which correspond the electrical depolarization of the ventricles, ideally leading to a mechanical contraction with ejection of blood formatting a pulse wave. Thus $z_a$ should exhibit a maximum at this shift if mechanical coupling is present, whereas we assume uncorrelated signals and therefore vanishing autocorrelation else. \\ 
Furthermore, we also use the second derivative $z_a''(n)$, approximated with finite stepsize $1/f_s$, to check the curvature of $z_a$ at $\tilde{n}_e$. In case of mechanical coupling the curvature of $z_a$ at the expected maximum $\tilde{n}_e$ the should be negative. We use $v_7=\overline{z''_a}:=\frac{1}{5} \sum_{i=-2}^2 z_a''(n)|_{n=\tilde{n_e}+i}$ to include also neighboring curvature values. The highly generic properties of these features and their independence from any hyperparameters makes them very stable in application. \\\
\noindent \textbf{3)} In case of arrhythmic ECG rhythms (e.g. atrial fibrillation, ventricular extrasystole), the methods relying on the autocorrelation fail, since there is no unique shift, for which the whole snippet is highly self-similar, even though characteristic acceleration patterns can still be associated to QRS-complexes (see Fig. \ref{fig:Arhy_Snippet}). To address this problem, we localize the QRS complexes similarly to the calculation of $n_P$ as described in \cite{Ayala:2014} by detecting their steep slopes: After applying a broad $4$-th order Butterworth bandpass filter with limiting frequencies $0.5$ and $30$ Hz to $e_n$,  we take the a $0.1$-second-rolling-mean of the square of first difference
$
\overline{(e^{(filt)}_{n+1}-e^{(filt)}_n)^2}_{0.1},
$
normalize it by dividing it through its maximum and find the local maxima which exceed $0.33$. The resulting candidates are further filtered by requiring different QRS-complexes to differ by at least $0.2$ s and taking the largest maximum for local maxima closer than $0.2$ s to each other as position of the QRS complex. Afterward, we cut potentially overlapping $0.48$ s-windows centered around the QRS-complexes from the ACC signal and compute a list of correlation values between each window with each other. If characteristic acceleration patterns can be linked to the QRS-complexes, the correlation between these windows should be large, whereas we expect them to be small else. Taking the $75$-th percentile of this list gives us a correlation feature $v_{8}$ for arrhythmic rhythms. Employing the $75$-th percentile instead of the mean balances well between correlation values that are high by coincidence and not taking windows with artifacts in the acceleration signal into account. These windowed correlation is identically computed for the ECG signal which yields $v_{9}$ and the ratio between the two correlations $v_{10}=v_{8}/v_{9}$ is used as a feature too.
In the case of ECG signals with no QRS-complexes, the algorithm still finds peaks with steep slopes and uses them for further processing. Since physiologically no accelerations by the apex beat should be detectable for these rhythms, the acceleration signals of the windows around those peaks should be uncorrelated to each other regardless of what peak the algorithm has detected.

For the ACC signal we further use features similar to the ones proposed in \cite{ashouri:2017}. We compute the ensemble average for all windowed ACC signals from above and take its root-mean-square, kurtosis, skewness, median, peak-to-peak-amplitude and peak-to-peak to root-mean-square ratio as time-domain features, as well as its band power in the following frequency bands ($0$-$3$Hz, $3$-$6$Hz, $6$-$9$Hz, $9$-$12$Hz, $12$-$15$Hz, $15$-$18$Hz)  and mean, standard deviation, kurtosis, skewness of the spectrogram, its maximum, the frequency, where the maximum appears and the spectral entropy of the 4-second ACC-signal \cite{solar:2017} as frequency domain features  ($v_{11},\dots v_{29}$).

For the ECG we further use the features $(v_{30},\dots v_{35}) =(P_\text{LEA}, L_\text{min}, b_S, n_P, P_\text{fib}, P_h)$ which were computed following \cite{Ayala:2014} as well as the \texttt{meanRR}, \texttt{VarRR}, \texttt{MeanPP}, \texttt{StdPP}, the mean and standard deviation of the QRS-width, \texttt{SlopeQRS}, \texttt{MSnorm}, \texttt{StdSnorm} from \cite{Elola:2020} and mean, and standard deviation of the absolute of the first difference of the signal, kurtosis of the square of the first difference , \texttt{AMSA} and \texttt{HfP}  (following \cite{Elola:2019}) which are $(v_{36},\dots v_{49})$. As proposed in the literature, we applied a broad $4$-th order Butterworth bandpassfilter ($0.8$-$30$) Hz to the ECG-signal before feature extraction.
These $49$ features and the label can be represented as $(\mathbf{x},y) \in \mathbb{R}^{49} \times \lbrace-1,1\rbrace$, where $y=+1$ is related to SC and $y=-1$ with its absence. We refer to $v_1$, $v_3$, $v_5-v_{8}$ and $v_{10}-v_{29}$ as 'ACC-features', and call the remaining features 'ECG-features'. Note that ECG-features use solely ECG data, whereas ACC-features employ ACC data or a combination of ACC and ECG.

\subsection{Learning/Training}
The total data set $(\mathbf{x},y)$ was randomly split into a training set $(\mathbf{x_\text{train}},y_\text{train})$ and a test set $(\mathbf{x_\text{test}},y_\text{test})$. The split was performed patient-wise so that all snippets from one patient belong entirely either to the training set or to the test set. The set sizes follow a roughly ratio of $3  : 1$, depending of the number of snippets per case in training and test set.  The training data were shifted to median $0$ and scaled to interquartile range (IQR) $1$, and the same transformation was applied to the test set. The training set was used to train a Gaussian kernel (rbf) Support Vector Machine (SVM) classification model with $l2$-regularization using the package SciKit-Learn \cite{scikit-learn} which employs LIBSVM package \cite{libsvm}. The optimal hyperparameters (kernel parameter $\gamma$ and regularization $C$) were chosen to maximize balanced accuracy on the following grid during $20$-fold cross-validation:  $\gamma \in \lbrace 10^k | k \in -5+\frac{n}{4}, n =0,\ldots,14  \rbrace $ and $C \in \lbrace 10^k | k \in -3+\frac{n}{3}, n =0,\ldots,15   \rbrace $. The performance of the model was assessed using balanced accuracy, sensitivity, specificity, Mathew's correlation coefficient (MCC), $F_{1}$-score of the 'SC'-class and the Area under the receiver-operator-characteristic-curve (AUC) on the test set.  \\
Since the results on the test set vary significantly depending on the chosen data-split the procedure above of splitting the data, training the algorithm in the training set (with cross-validation), and evaluating on the test set was repeated $50$ times and evaluated by reporting mean and $95\%$-confidence interval (CI) of the distribution of performance measures for the different data splits. \\
We trained models using only ECG-features too, to compare with our proposed classifier and other existing ones.  \\
The code for the classifier and the scaler using all features together with 5 exemplary cases from the test set is made publicly available in \cite{CPRDAT}.

\section{Results}
\subsection{Data base}

From the total sample of $871$ cases, $55$ recordings were excluded from further analysis since the files were corrupted, for $92$ cases no conclusive annotation was feasible or the annotation did not agree with the registry entries regarding the occurrence of at least one ROSC, and in $158$ cases no simultaneous ECG, ACC, and capnography signals were available. $127$ of the remaining cases lack interruptions in CPR with appropriate length or lack signal overlap before the end of the last chest compression period. So a total of $439$ cases were included, yielding $34903$ snippets. During prefiltering as described in \ref{ssec:preprocessing}, $9579$ ($189$ ASY, $626$ VF, $1561$ VT, $7203$ ORG) snippets got omitted, so that the final data base consists of $25324$ Snippets from $422$ cases. These cases cover $250.3$ h of defibrillator recordings and $173.6$ h of cardiac arrest. Chest compressions were interrupted $7528$ times covering $133.6$ h with a median (IQR) length of $5.35~ (2.80,14.3)$s. $6099$ of those interuptions were pauses without spontaneous circulation covering $14.3$ h, their median (IQR) length was $4.52~(2.43,10.2)$ s. The $1429$ interruptions with spontaneous circulation present covered a period of $119.3$ h in total, their median (IQR) length was $96.6~ (11.6,418)$ s. Further information about the data set can be found in Table \ref{tab:data} and an Utstein-style data sheet is provided in the supplemental material. 
\begin{table}
\centering
\caption{Overview over total data set. Number of snippets per circulatory state and ECG rhythm.}
\label{tab:data}
 \begin{tabular}{|l|c|}
 \hline 
 \textbf{Total number of Cases} & 422 \\
  \hline 

 \textbf{Cases with ROSC} & 199   \\ 
 \hline 
  \textbf{Snippets} & 25324  \\
    \hline 

   \textbf{Circulatory state} &     \\ 
  Spontaneous circulation (SC)  & 14993  \\
  Cardiac Arrest (AR) & 10331  \\
    \hline 

     \textbf{Rhythm} &\\ 
    Asystole (ASY) & 945   \\ 
    Ventricular Fibrillation (VF) & 1953  \\
    Ventricular Tachycardia (VT) & 1834 \\
    Organized Rhythm (ORG) & 20592 \\
 \hline 
 \end{tabular} 
\end{table}

\subsection{Performance}

\begin{figure}
\centering
\includegraphics[trim=1.5cm 0.0cm 1.5cm 1.2cm,clip,width=.5 \textwidth]{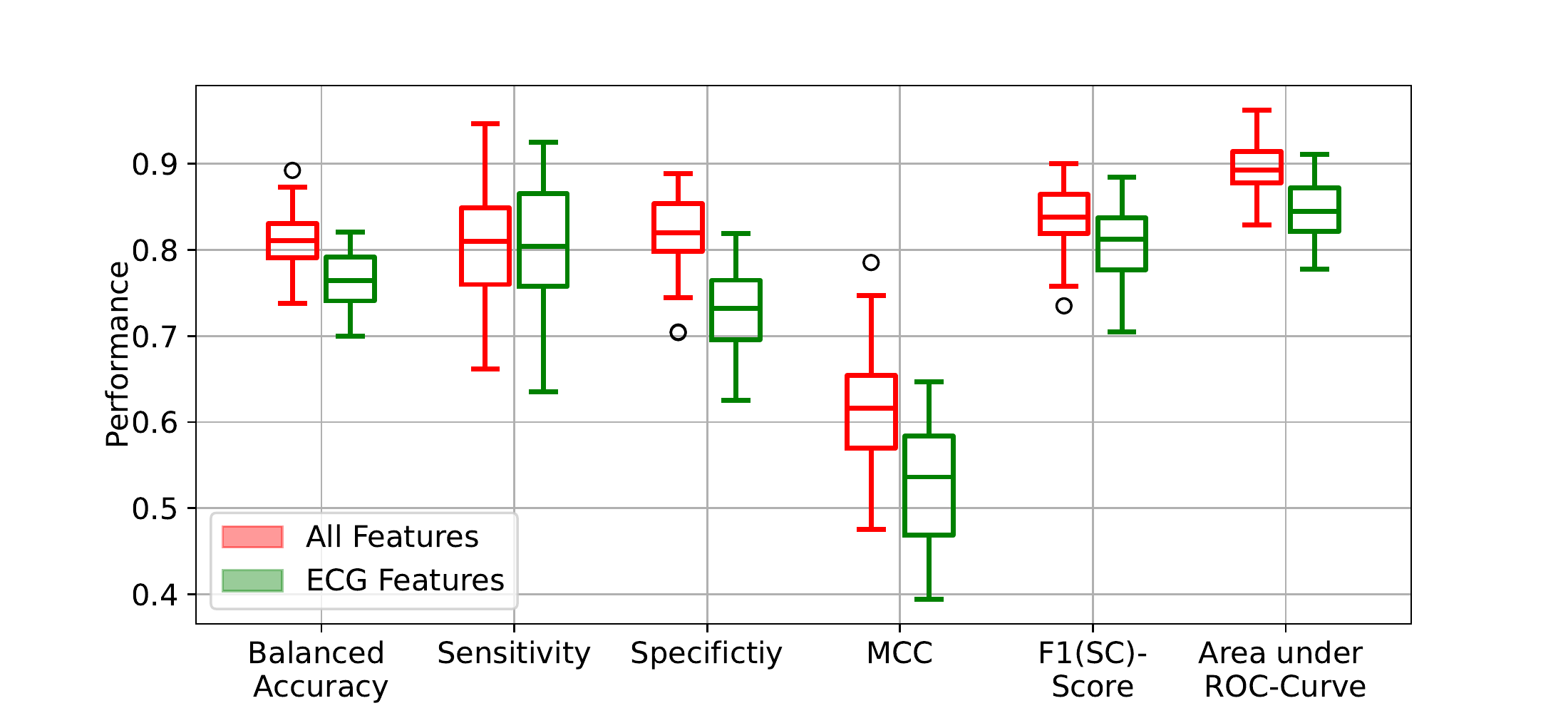}
\caption{Performance of models using all features or only ECG features evaluated for different data splittings.}
\label{fig:nested_cv_results}
\end{figure}

On 50 different data splits the classifier exhibits a balanced accuracy of $0.812~(0.749,0.875)$, a sensitivity of $0.806~(0.68,0.932)$, a specificity of $0.818~(0.735,0.901)$ and a MCC of $0.614~(0.479,0.748)$.
The performance using only ECG features, and the mean ($95\%$-CI) gain in performance per data split are given in Table \ref{tab:nested_CV_results}. The distribution of different performance measures on the data splits are given in Fig. \ref{fig:nested_cv_results}. While sensitivity does not change significantly by adding accelerometer data, specifictiy increases by around $9\%$. Thus, snippets which would have been classified as 'SC' by ECG only are now correctly classified as 'AR'.  

\begin{table*}
\caption{Mean ($95\%$-confidence interval) values of performance and performance gain per data split when adding ACC-features. }
\label{tab:nested_CV_results}
\setlength{\tabcolsep}{3pt}
\begin{tabular}{|c|c|c|c|}
\hline 
\textbf{Performance} & \textbf{All Features} & \textbf{ECG Features} & \textbf{Performance gain} \\ 
\hline 
Balanced Accuracy & $0.812$  & $0.765$ & $0.047$ \\ 
  & $(0.749,0.875)$ & $(0.705,0.826)$ & $(-0.006,0.0998)$ \\
\hline 
Sensitivity & $0.806$  & $0.802$ & $0.004$ \\ 
  & $(0.68,0.932)$ & $(0.661,0.944)$ & $(-0.117,0.125)$ \\
\hline 
Specificity & $0.818$   &  $0.728$ & $0.090$ \\ 
  & $(0.735,0.901)$ & $(0.632,0.824)$ & $(0.022,0.157)$ \\
\hline 
MCC & $0.614$ & $0.528$ & $0.085$ \\ 
  & $(0.479,0.748)$ & $(0.394,0.663)$ & $(-0.027,0.197)$ \\
\hline 
$F_{1}$ & $0.835$ & $0.806$ & $0.028$ \\ 
  & $(0.758,0.911)$ & $(0.72,0.893)$ & $(-0.043,0.100)$ \\
\hline 
Area under Receiver-  & $0.896$ & $0.846$ & $0.050$ \\ 
operator-characteristic  & $(0.843,0.948)$ & $(0.782,0.911)$ & $(0.003,0.096)$ \\
 curve (AUC) & & & \\
\hline 
\end{tabular} 
\end{table*}

\begin{figure}
\centering
\includegraphics[trim=1.3cm 0.6cm 1.8cm 1.4cm,clip,width=.5 \textwidth]{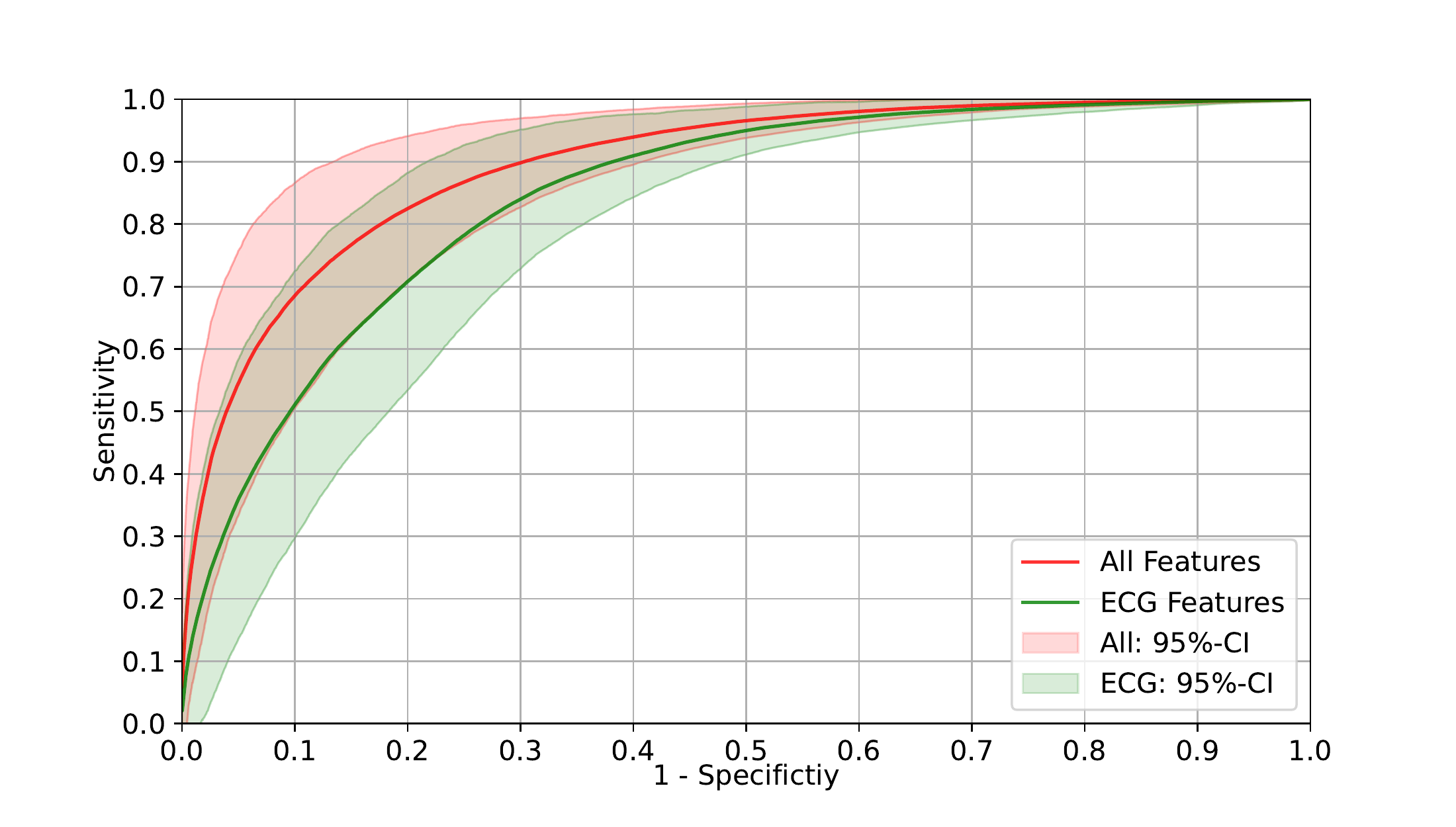}
\caption{Mean and $95\%$-confidence interval of the ROC-curve for both models with all features and ECG features only respectively.}
\label{fig:roc_curves}
\end{figure}

Mean and $95\%$-CI of the receiver-operator-characteristic (ROC)-curve are shown in Fig. \ref{fig:roc_curves}.

We further aim to compare our algorithm directly to the rhythm classifier by Rad et al \cite{Rad:2017}. This can be achieved by summarizing the rhythms ASY, PEA and VF to our AR-class whereas PR belongs to our SC-class. For VT rhythms, one can not decide without further knowledge, whether this rhythm generates pulse. Thus, we omit VT rhythms in the test sets, and evaluate on the remaining test sets. The results, shown in Table \ref{tab:Rad_alg}, exhibit similar balanced accuracy for our ECG-feature-employing algorithm and the algorithm by Rad et al \cite{Rad:2017}. The larger deviations in sensitivity and specificity might be caused by a different proportion of 'SC' and 'AR'-snippets in the training set of the algorithm by \cite{Rad:2017}, leading to a different best operating point on the ROC-curve. 

\begin{table*}
\centering
\caption{Comparison to Algorithm by Rad et al \cite{Rad:2017} on test set without Ventricular Tachycardia (VT) rhythms.}
\label{tab:Rad_alg}
\setlength{\tabcolsep}{3pt}
\begin{tabular}{|c|c|c|c|}
\hline 
\textbf{Performance} & \textbf{All Features} & \textbf{ECG Features} & \textbf{Rad et al \cite{Rad:2017}} \\ 
\hline 
Balanced Accuracy & $0.822 $ & $0.775 $ & $ 0.761$ \\ 
& $(0.761,0.883)$ & $(0.709,0.840)$ & $(0.687,0.835)$ \\
\hline 
Sensitivity & $0.814 $ & $0.806 $ & $0.746 $ \\ 
& $(0.687,0.942)$ & $(0.656,0.955)$ & $(0.605,0.887)$ \\
\hline 
Specificity & $0.830 $ & $0.744 $ & $ 0.777 $ \\ 
& $(0.747,0.912)$ & $(0.645,0.842)$ & $(0.706,0.847)$ \\
\hline 
MCC & $ 0.635$ & $ 0.548$ & $ 0.513$ \\ 
& $(0.504,0.766)$ & $(0.403,0.692)$ & $(0.363,0.662)$ \\
\hline 
$F_{1}$ & $0.841 $ & $0.810 $ & $ 0.782$ \\ 
& $(0.766,0.917)$ & $(0.721,0.899)$ & $(0.684,0.881)$ \\
\hline 
\end{tabular} 
\end{table*}
Moreover, we are interested in the performance of our algorithm on organized rhythms, since for these the circulatory state can not inferred easily by ECG only, and this case is also widely discussed on the literature (e.g. \cite{Elola:2020}). The results are shown in Table \ref{tab:PR/PEA}. Compared to the performance on the entire data set, specificity drops about $4.6\%$ when using all features, but around $7.2\%$ when using ECG features, whereas sensitivities sligthly increase. This indicates improved PEA detection when employing ACC data.   
\begin{table*}
\centering
\caption{Mean ($95\%$-confidence interval) values of performance of different algorithms for pure Pulsative rhythms (PR) vs. Pulseless electric activity (PEA) discrimination.}
\label{tab:PR/PEA}
\setlength{\tabcolsep}{3pt}
\begin{tabular}{|c|c|c|c|}
\hline 
\textbf{Performance} & \textbf{All Features} & \textbf{ECG Features} & \textbf{Rad et al \cite{Rad:2017}} \\ 
\hline 
Balanced Accuracy & $0.793 $ & $0.732 $ & $ 0.729$ \\ 
& $(0.731,0.855)$ & $(0.669,0.796)$ & $(0.656,0.802)$ \\

\hline 
Sensitivity & $0.814 $ & $0.809 $ & $0.761 $ \\ 
& $(0.685,0.943)$ & $(0.661,0.956)$ & $(0.629,0.894)$ \\

\hline 
Specificity & $0.772 $ & $0.656 $ & $ 0.696 $ \\ 
& $(0.661,0.884)$ & $(0.53,0.783)$ & $(0.601,0.791)$ \\

\hline 
MCC & $ 0.572$ & $ 0.465$ & $ 0.443$ \\ 
& $(0.434,0.709)$ & $(0.318,0.611)$ & $(0.292,0.595)$ \\

\hline 
$F_{1}$ & $0.841 $ & $0.811 $ & $ 0.791$ \\ 
& $(0.765,0.917)$ & $(0.723,0.9)$ & $(0.696,0.886)$ \\

\hline 
\end{tabular} 
\end{table*}

\subsection{Feature importance}
\begin{figure}
\centering
\includegraphics[trim=1.4cm 0.2cm 1.8cm 1.3cm,clip,width=.5 \textwidth]{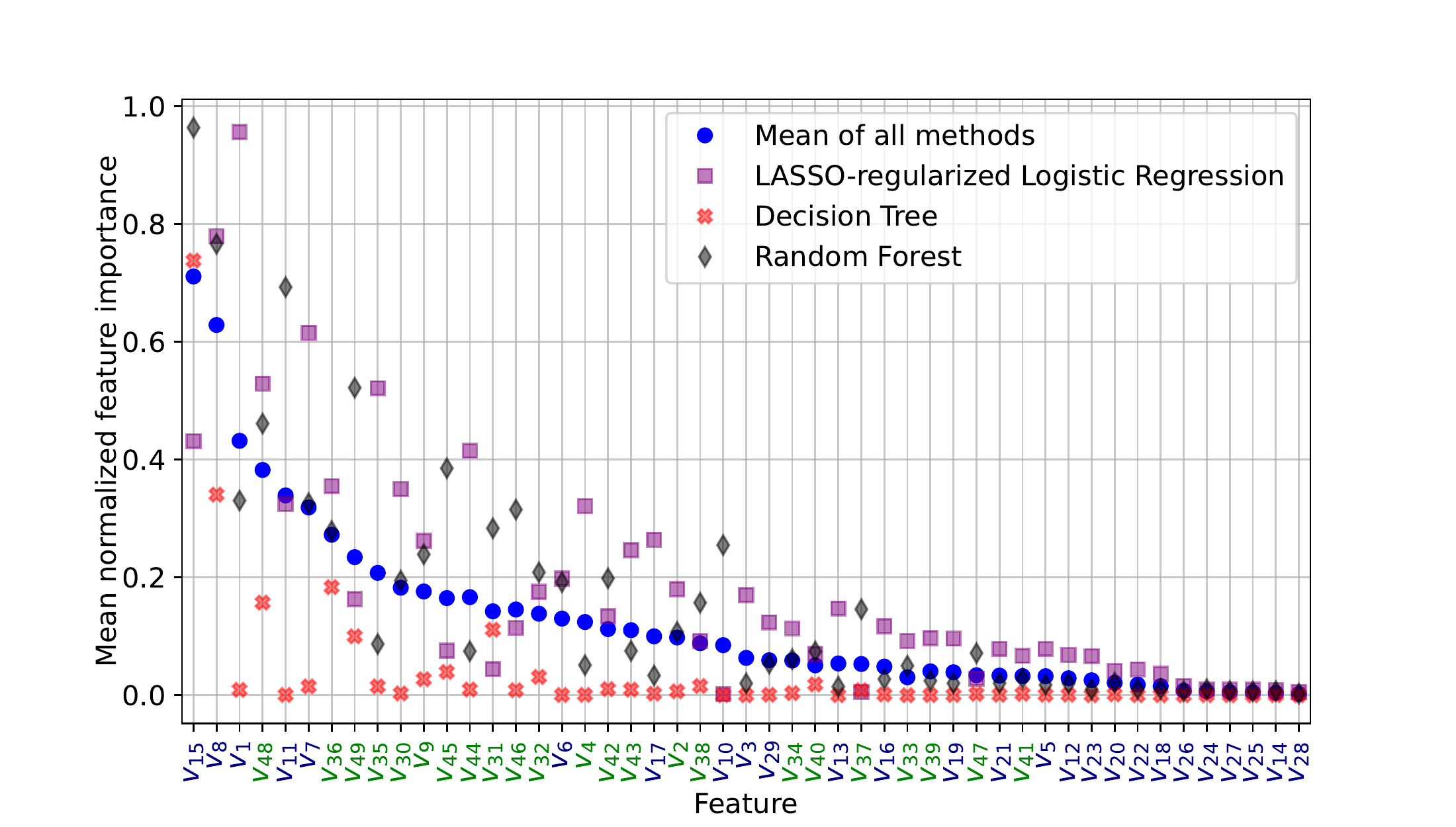}
\caption{Features importance of Lasso-regularized Logistic Regression, Decision Trees and Random Forests and their mean. ACC/ECG-features are shown in blue/green font color respectively.}
\label{fig:feat_importance}
\end{figure}

To analyze the importance of the newly developed features, we further trained three classifiers based on algorithms where feature importance is explicit (Lasso-regularized logistic regression, Decision Tree, Random Forests) on the 50 data splits, and analyzed their mean feature importance. The results in Fig. \ref{fig:feat_importance} show that the most important features are ACC-features, illustrating the significance of the ACC signal for an improved pulse/no-pulse decision. A detailed list with all features and their importances can be found in the supplemental material. Training an exemplary kernelized SVM with only the 10 best-ranked features yields a classifier with further improved performance: (balanced accuracy: $0.820, (0.760,0.878)$, sensitivity: $0.821, (0.700,0.943)$, specificity: $0.820, (0.750,0.885)$, MCC: $0.631, (0.510,0.756)$, $F_1$: $0.844, (0.770,0.917)$). This indicates that a tailored optimal feature selection would probably even further increase the performance and could be the content of future research. The feature importance analysis using only ECG features and the performance of a classifier with only the best 10 ECG features can be found in the supplemental material. 

\subsection{Case Studies}
\label{ssec:Case_Studies}
\begin{figure*}
\centering
\begin{subfigure}{1 \textwidth}
	\includegraphics[trim=3cm 0.5cm 4cm 0.7cm,clip,width=\textwidth]{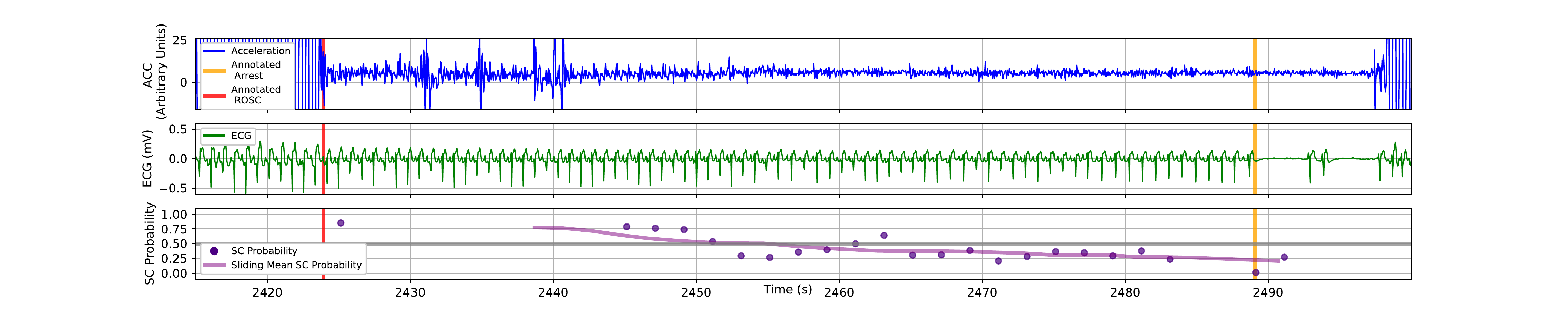}
	\vspace{-0.6cm}
	\caption{Ambigous Arrest annotation. }
\label{fig:Arrest}
\end{subfigure}
\begin{subfigure}{1 \textwidth}
	\includegraphics[trim=3cm 0.5cm 4cm 0.7cm,clip,width=\textwidth]{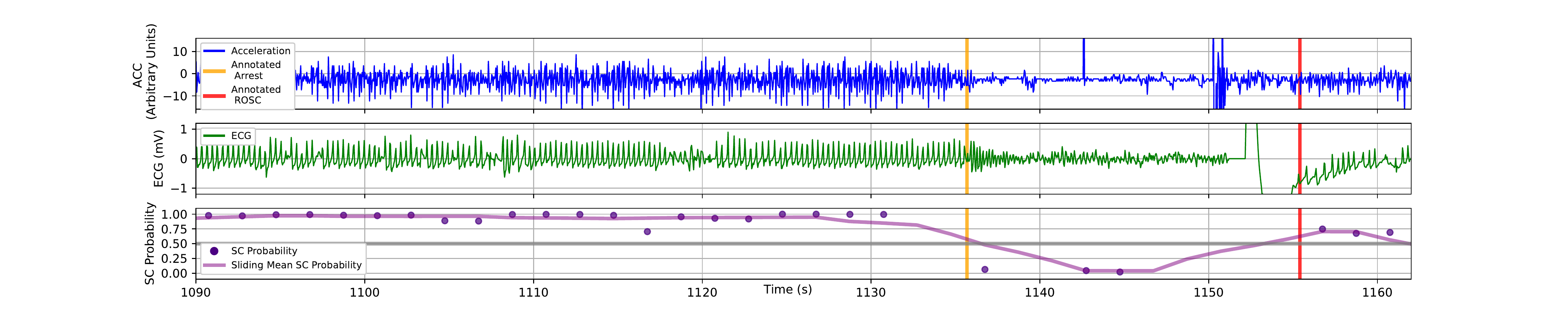}
	\vspace{-0.6cm}
	\caption{Arrest with following successful shock. }
\label{fig:Shock}
\end{subfigure}
\vspace*{-0.1cm}
\caption{Two exemplary situations with predicted spontaneous circulation probabilities over time. Acceleration data (blue), ECG data (green) and probabilities (purple scatters), and a sliding mean of probabilities (purple solid line) are given. Additionally, the labels by the manual, retrospective annotation and the time of the shock (yellow diamonds) are shown. Both cases are taken from the test set.}
\label{fig:Timeline}
\end{figure*}

The performance of the algorithm can be illustrated by plotting the predictions and their probability of successive snippets over time.  In Fig. \ref{fig:Timeline}, two exemplary situations are shown.  In Fig. \ref{fig:Arrest}, the algorithm predicts the probability of spontaneous circulation to be below $50\%$ at approximately $25$ s earlier than retrospectively annotated, although the ECG rhythm shows no significant change there. Only the peaks of the acceleration signal decrease. In Fig. \ref{fig:Shock} the arrest time can be defined unambiguously by the abrupt rhythm change in ECG. Correspondingly, the algorithm predicts AR subsequently. After the shock, the organized rhythm starts again, and the algorithm predicts SC with high probability. \\
The performance of the classifier on 5 different cases is furthermore illustrated in \cite{CPRDAT}.

\section{Discussion}
To the best of our knowledge, this study is the first to employ accelerometry data to classify the circulatory state in cases of out-of-hospital cardiac arrest treatment. 
In contrast to other studies investigating seismocardiography in porcine models \cite{Wei:2017} or under lab conditions \cite{Lee:2021}, this work employs retrospective analysis of prospectively collected real-world data from defibrillator recordings, leading to conceptual advantages on the one hand, but causing difficulties on the other hand too. \\
Using real-world recordings demonstrates the applicability of accelerometry to assess the circulatory state also on noisy signals from the field. Although the main purpose of the defibrillator models' accelerometer used in this study is the assessment of CPR quality delivered by the EMS, the apex beat is still detectable by these devices. However, since acceleration amplitudes by the apex beat are comparatively small, discretization artifacts are already visible in the recordings, inevitable noise and artifacts from the transport or medical treatment cover the signal easily and the signal quality is highly dependent on the exact placement of the sensor on the patient's chest  \cite{Taebi:2019}, analysis is complicated considerably. Furthermore, the defibrillator recordings from cardiac arrest patients contain data from ischemic hearts where inotropy might be reduced, wall motion abnormalities may be present and subsequently signal quality might be worse than for healthy patients.
Moreover, the measuring range of the accelerometer is adjusted to accelerations occurring in CPR. Using devices with measuring ranges adapted to the expected signal strength and optimizing the position of the accelerometer on the chest could improve signal quality considerably. Even though the best configuration requires further investigation, the respectable performance of the algorithm demonstrates its applicability in the current configuration too.
It is also worth noting that, as alternative to feature-based classifiers, we have also experimented with neural-network-based methods operating directly on the snippets, but in all of our experiments, feature-based classifiers performed best. Moreover, the best performance of our algorithm was obtained with rather high regularization parameters (geometric mean over all data splits: $\gamma=0.00072$, $C=2.09$), which indicates that, in the presence of many features, the proposed regularization is important to prevent overfitting.\\
One key difficulty of our data collection is the determination of a reliable ground truth. Retrospective annotation of the circulatory state is inevitably imprecise for several reasons. First, the annotators have to assess the circulatory state based on limited information provided by recordings and documentation so that a concluding annotation might not be feasible. We tried to encounter this problem by assuming faultless treatment by the EMS and exclusion of obtuse intervals during the annotation process.
Second, the dichotomous annotation design does not take account of periarrest situations like pseudo-PEA, where cardiac contraction is present but insufficient. These situations can not be classified unambiguously within this framework. Since the heart is also palpating in pseudo-PEAs our algorithm could still be capable of detecting the apex beat, even if cardiac output is insufficient. A better assessment of periarrest rhythms requires more intricate measurements of cardiac output like echocardiography or invasive arterial blood pressure measurement, which are not present in our data set. Employing such measurements could further investigate the question of whether accelerometry is capable of assessing the circulatory state not only qualitatively but also quantifying cardiac contractility which has to remain unanswered due to a lack of appropriate data. \\
Finally, only one year was annotated by two independent physicians, where dissenting annotations were resolved in consensus. The rest of the data were labeled by a single physician. In addition, all data were collected from one model (ZOLL X-Series) in this study. However, as long as the accelerometer records with at least similar resolution and appropriate sample rate and its position is adequate to record the apex beat, the same working principle should be applicable to other models too.   \\
In contrast to other circulation classification algorithms proposed in the literature, this algorithm operates on a more general data set containing not only organized (PR and PEA), but also any other ECG rhythms where mechanical decoupling can be assessed directly from ECG data (e.g. VF  or ASY). Nevertheless, adding information from the accelerometer increases classification performance considerably, enhancing specificity around $9\%$ on average for all data splits. This means that snippets were classified as AR correctly, which would have been classified as SC from ECG only. Furthermore, the acceptance of any ECG and acceleration signal pair as input allows for an application in more general situations with pulse generating arrythmias present. \\
However, using the rhythm classifier by Rad et al \cite{Rad:2017}, allows for a comparison of our results to the literature by evaluating our algorithm on certain rhythm types on the test sets only, although direct comparison has to be carried out with caution, since evaluating on different data sets can cause a large variation in performance. Nevertheless, using accelerometry data increases specificity on average around  $11.5\%$ for PR/PEA detection, emphasizing the significant role of accelerometry for PR/PEA classification, which is superior to the specificity gain of $9\%$ found by  Elola et al. \cite{Elola:2020} when adding thoracic impedance data to ECG data.
Moreover, compared to the usage of thoracic impedance acclerometry exhibits a big advantage:  Whereas ventilation and respiration induce large amplitude signals in thoracic impedance \cite{Elola:2019b}, requiring advanced filtering techniques and  potentially complicating circulatory state assessment, the slow movement of the chest wall during ventilation or respiration is hardly detectable in the acceleration signal and no filtering techniques are required. Furthermore, combining ECG, ACC, thoracic impedance and potentially CO2 data could enhance performance further. \\
Due to high variability in the data set, performance varies a lot from one data split to another, leading to broad confidence intervals for the performance measures. It is worth noting,  however, that AUC and specificity increases at every data split by adding ACC-features whereas in $48$ from $50$ data splits, balanced accuracy and MCC are increasing too. The decrease in the two remaining data splits is very small, about $1\%$ and $0.4\%$ in balanced accuracy. Thus, we interpret the gain in performance as relevant despite the broad confidence intervals, since performance increases in $96 \%$ of all data splits. \\
The porcine model by Wei et al. \cite{Wei:2017} employing accelerometry exhibited a sensitivity of $93.6 \%$ and a specificity of $97.5 \%$ to discriminate PEA from PR under lab conditions, which hardly can be compared to real world human data. They proposed the usage of a single feature in their work which was not used as an input due to inevitable noise in our data.  \\
Regarding applicability in real-world cardiac arrest treatment, one could additionally exploit the chronological order of the snippets. So far, the algorithm does not include the snippets' temporal information. The prediction of a circulatory state of a single snippet does not take the state of the previous or next snippet into account. However, in real-world, consecutive snippets will mostly be related to each other since circulatory state transitions do not take place every few seconds normally, allowing for an identification of single misclassifications. \\
Employing temporal information is also relevant to assess the medical practicality of this algorithm in the field. During cardiac arrest treatment, a quick and reliable recognition of circulatory state changes is necessary to reduce no-flow time. The capability of the algorithm to detect ROSC and rearrest events within a short time interval on unknown cases needs further investigation in order to evaluate its usefulness as a support tool. 
 \\
Besides the usage of the algorithm in the field during cardiac arrest treatment, another potential application regards retrospective analysis. Currently, manual annotation is necessary to evaluate key CPR quality metrics like chest compression fraction. While recent works \cite{kramerjohansen:2007,Orlob:2021,Kern:2021, gupta:2021}  proposed methods how to compute the time with ongoing chest compressions automatically, the length of the cardiac arrest intervals still needs manual annotation. Our proposed algorithm could facilitate manual annotation by suggesting appropriate intervals, which physicians only need to approve instead of single-handed analyzing the whole case in detail. This assistance could allow for broader application of CPR quality metrics in general registries \cite{JAUREGUIBEITIA:2021}.
\section{Conclusion}

This study presented a machine learning algorithm using ECG and ACC data to determine the circulatory state of cardiac arrest patients. We explained the development of new features assessing the interdependence of ACC and ECG signals, and the medical rational for them. We showed that accelerometry provides a highly relevant input for a circulation detection algorithm in cardiac arrest patient and an accelerometry-based algorithms could serve as a valuable medical decision tool to provide optimal treatment for cardiac arrest patients. 

\section*{Acknowledgment}
We thank the Reviewers for their valuable remarks and we thank the authors of \cite{Rad:2017}, in particular A.B. Rad and T. Eftest\o{}l, for providing us with a pretrained version of their algorithm for rhythm analysis.

\bibliographystyle{unsrt}
\onecolumn
\bibliography{refs}

\begin{thebibliography}{10}

\bibitem{Grasner:2021}
Jan-Thorsten Gräsner, Johan Herlitz, Ingvild~B.M. Tjelmeland, Jan Wnent,
  Siobhan Masterson, Gisela Lilja, Berthold Bein, Bernd~W. Böttiger, Fernando
  Rosell-Ortiz, Jerry~P Nolan, Leo Bossaert, and Gavin~D. Perkins.
\newblock European resuscitation council guidelines 2021: Epidemiology of
  cardiac arrest in europe.
\newblock {\em Resuscitation}, 161:61--79, 2021.

\bibitem{ochoa:1998}
F~Javier Ochoa, E~Ramalle-Gomara, JM~Carpintero, A~Garc{\i}a, and I~Saralegui.
\newblock Competence of health professionals to check the carotid pulse.
\newblock {\em Resuscitation}, 37(3):173--175, 1998.

\bibitem{eberle:1996}
B.~Eberle, W.F. Dick, T.~Schneider, G.~Wisser, S.~Doetsch, and I.~Tzanova.
\newblock Checking the carotid pulse check: diagnostic accuracy of first
  responders in patients with and without a pulse.
\newblock {\em Resuscitation}, 33(2):107--116, 1996.

\bibitem{christenson:2009}
Jim Christenson, Douglas Andrusiek, Siobhan Everson-Stewart, Peter Kudenchuk,
  David Hostler, Judy Powell, Clifton~W Callaway, Dan Bishop, Christian
  Vaillancourt, Dan Davis, et~al.
\newblock Chest compression fraction determines survival in patients with
  out-of-hospital ventricular fibrillation.
\newblock {\em Circulation}, 120(13):1241--1247, 2009.

\bibitem{Soar:2021}
Jasmeet Soar, Bernd~W. Böttiger, Pierre Carli, Keith Couper, Charles~D.
  Deakin, Therese Djärv, Carsten Lott, Theresa Olasveengen, Peter Paal,
  Tommaso Pellis, Gavin~D. Perkins, Claudio Sandroni, and Jerry~P. Nolan.
\newblock European resuscitation council guidelines 2021: Adult advanced life
  support.
\newblock {\em Resuscitation}, 161:115--151, 2021.
\newblock European Resuscitation Council Guidelines for Resuscitation 2021.

\bibitem{Elola:2019}
Andoni Elola, Elisabete Aramendi, Unai Irusta, Javier Del~Ser, Erik Alonso, and
  Mohamud Daya.
\newblock Ecg-based pulse detection during cardiac arrest using random forest
  classifier.
\newblock {\em Medical \& biological engineering \& computing}, 57(2):453--462,
  2019.

\bibitem{Elola:2019a}
Andoni Elola, Elisabete Aramendi, Unai Irusta, Artzai Picón, Erik Alonso,
  Pamela Owens, and Ahamed Idris.
\newblock Deep neural networks for ecg-based pulse detection during
  out-of-hospital cardiac arrest.
\newblock {\em Entropy}, 21(3), 2019.

\bibitem{Kwok:2020}
Heemun Kwok, Jason Coult, Jennifer Blackwood, Shiv Bhandari, Peter Kudenchuk,
  and Thomas Rea.
\newblock Electrocardiogram-based pulse prediction during cardiopulmonary
  resuscitation.
\newblock {\em Resuscitation}, 147:104--111, 2020.

\bibitem{Cromie:2008}
Nick~Alexander Cromie, John~Desmond Allen, Colin Turner, John~McC Anderson, and
  A~A~Jennifer Adgey.
\newblock The impedance cardiogram recorded through two
  electrocardiogram/defibrillator pads as a determinant of cardiac arrest
  during experimental studies.
\newblock {\em Critical care medicine}, 36(5):1578--1584, 2008.

\bibitem{Elola:2019b}
Andoni Elola, Elisabete Aramendi, Unai Irusta, Artzai Pic{\'o}n, Erik Alonso,
  Iraia Isasi, and Ahamed Idris.
\newblock Convolutional recurrent neural networks to characterize the
  circulation component in the thoracic impedance during out-of-hospital
  cardiac arrest.
\newblock In {\em 2019 41st Annual International Conference of the IEEE
  Engineering in Medicine and Biology Society (EMBC)}, pages 1921--1925. IEEE,
  2019.

\bibitem{Risdal:2008}
Martin Risdal, Sven~Ole Aase, Jo~Kramer-Johansen, and Trygve Eftestol.
\newblock Automatic identification of return of spontaneous circulation during
  cardiopulmonary resuscitation.
\newblock {\em IEEE Transactions on Biomedical Engineering}, 55(1):60--68,
  2008.

\bibitem{Alonso:2016}
Erik Alonso, Elisabete Aramendi, Mohamud Daya, Unai Irusta, Beatriz Chicote,
  James~K. Russell, and Larisa~G. Tereshchenko.
\newblock Circulation detection using the electrocardiogram and the thoracic
  impedance acquired by defibrillation pads.
\newblock {\em Resuscitation}, 99:56--62, 2016.

\bibitem{Ruiz:2018}
Jesus~M. Ruiz, Sofía {Ruiz de Gauna}, Digna~M. González-Otero, Purificación
  Saiz, J.~Julio Gutiérrez, Jose~F. Veintemillas, Jose~M. Bastida, and Daniel
  Alonso.
\newblock Circulation assessment by automated external defibrillators during
  cardiopulmonary resuscitation.
\newblock {\em Resuscitation}, 128:158--163, 2018.

\bibitem{Elola:2019c}
Andoni Elola, Elisabete Aramendi, Unai Irusta, Erik Alonso, Yuanzheng Lu,
  Mary~P Chang, Pamela Owens, and Ahamed~H Idris.
\newblock Capnography: A support tool for the detection of return of
  spontaneous circulation in out-of-hospital cardiac arrest.
\newblock {\em Resuscitation}, 142:153--161, 2019.

\bibitem{Elola:2020}
Andoni Elola, Elisabete Aramendi, Unai Irusta, Per~Olav Berve, and Lars Wik.
\newblock Multimodal algorithms for the classification of circulation states
  during out-of-hospital cardiac arrest.
\newblock {\em IEEE Transactions on Biomedical Engineering}, 68(6):1913--1922,
  2021.

\bibitem{Wijshoff:2015}
Ralph W. C. G.~R. Wijshoff, Antoine M. T.~M. van Asten, Wouter~H. Peeters, Rick
  Bezemer, Gerrit~Jan Noordergraaf, Massimo Mischi, and Ronald~M. Aarts.
\newblock Photoplethysmography-based algorithm for detection of cardiogenic
  output during cardiopulmonary resuscitation.
\newblock {\em IEEE Transactions on Biomedical Engineering}, 62(3):909--921,
  2015.

\bibitem{cohen:2022}
Allison~L. Cohen, Timmy Li, Lance~B. Becker, Casey Owens, Neha Singh, Allen
  Gold, Mathew~J. Nelson, Daniel Jafari, Ghania Haddad, Alexander~V. Nello,
  Daniel~M. Rolston, Cristina Sison, and Martin~L. Lesser.
\newblock Femoral artery doppler ultrasound is more accurate than manual
  palpation for pulse detection in cardiac arrest.
\newblock {\em Resuscitation}, 173:156--165, 2022.

\bibitem{Taebi:2019}
Amirtah{\`a} Taebi, Brian~E Solar, Andrew~J Bomar, Richard~H Sandler, and
  Hansen~A Mansy.
\newblock Recent advances in seismocardiography.
\newblock {\em Vibration}, 2(1):64--86, 2019.

\bibitem{Wei:2017}
Liang Wei, Gang Chen, Zhengfei Yang, Tao Yu, Weilun Quan, and Yongqin Li.
\newblock Detection of spontaneous pulse using the acceleration signals
  acquired from cpr feedback sensor in a porcine model of cardiac arrest.
\newblock {\em PLOS ONE}, 12(12):1--11, 12 2017.

\bibitem{Lee:2021}
Hyoung~Youn Lee, Yong~Hun Jung, Kyung~Woon Jeung, Dong~Hun Lee, Byung~Kook Lee,
  Geuk~Young Jang, Tong~In Oh, Najmiddin Mamadjonov, and Tag Heo.
\newblock Discrimination between the presence and absence of spontaneous
  circulation using smartphone seismocardiography: A preliminary investigation.
\newblock {\em Resuscitation}, 166:66--73, 2021.

\bibitem{jupyter}
Thomas Kluyver, Benjamin Ragan-Kelley, Fernando P{\'e}rez, Brian Granger,
  Matthias Bussonnier, Jonathan Frederic, Kyle Kelley, Jessica Hamrick, Jason
  Grout, Sylvain Corlay, Paul Ivanov, Dami{\'a}n Avila, Safia Abdalla, and
  Carol Willing.
\newblock Jupyter notebooks -- a publishing format for reproducible
  computational workflows.
\newblock In F.~Loizides and B.~Schmidt, editors, {\em Positioning and Power in
  Academic Publishing: Players, Agents and Agendas}, pages 87 -- 90. IOS Press,
  2016.

\bibitem{kramerjohansen:2007}
Jo~Kramer-Johansen, Dana~P Edelson, Heidrun Losert, Klemens K{\"o}hler, and
  Benjamin~S Abella.
\newblock Uniform reporting of measured quality of cardiopulmonary
  resuscitation (cpr).
\newblock {\em Resuscitation}, 74(3):406--417, 2007.

\bibitem{Orlob:2021}
Simon Orlob, Wolfgang~J. Kern, Birgitt Alpers, Michael Schörghuber, Andreas
  Bohn, Martin Holler, Jan-Thorsten Gräsner, and Jan Wnent.
\newblock Chest compression fraction calculation: A new, automated, robust
  method to identify periods of chest compressions from defibrillator data –
  tested in zoll x series.
\newblock {\em Resuscitation}, 2022.

\bibitem{Kern:2021}
Wolfgang~J. Kern, Simon Orlob, Birgitt Alpers, Michael Schörghuber, Andreas
  Bohn, Martin Holler, Jan-Thorsten Gräsner, and Jan Wnent.
\newblock A sliding-window based algorithm to determine the presence of chest
  compressions from acceleration data.
\newblock {\em Data in Brief}, page 107973, 2022.

\bibitem{Rad:2017}
Ali~Bahrami Rad, Trygve Eftest{\o}l, Kjersti Engan, Unai Irusta, Jan~Terje
  Kval{\o}y, Jo~Kramer-Johansen, Lars Wik, and Aggelos~K Katsaggelos.
\newblock Ecg-based classification of resuscitation cardiac rhythms for
  retrospective data analysis.
\newblock {\em IEEE Transactions on Biomedical Engineering}, 64(10):2411--2418,
  2017.

\bibitem{Ayala:2014}
Unai Ayala, U~Irusta, J~Ruiz, T~Eftest{\o}l, J~Kramer-Johansen,
  F~Alonso-Atienza, E~Alonso, and D~Gonz{\'a}lez-Otero.
\newblock A reliable method for rhythm analysis during cardiopulmonary
  resuscitation.
\newblock {\em BioMed research international}, 2014, 2014.

\bibitem{ashouri:2017}
Hazar Ashouri and Omer~T Inan.
\newblock Automatic detection of seismocardiogram sensor misplacement for
  robust pre-ejection period estimation in unsupervised settings.
\newblock {\em IEEE sensors journal}, 17(12):3805--3813, 2017.

\bibitem{solar:2017}
Brian~E Solar, Amirtaha Taebi, and Hansen~A Mansy.
\newblock Classification of seismocardiographic cycles into lung volume phases.
\newblock In {\em 2017 IEEE Signal Processing in Medicine and Biology Symposium
  (SPMB)}, pages 1--2. IEEE, 2017.

\bibitem{scikit-learn}
F.~Pedregosa, G.~Varoquaux, A.~Gramfort, V.~Michel, B.~Thirion, O.~Grisel,
  M.~Blondel, P.~Prettenhofer, R.~Weiss, V.~Dubourg, J.~Vanderplas, A.~Passos,
  D.~Cournapeau, M.~Brucher, M.~Perrot, and E.~Duchesnay.
\newblock Scikit-learn: Machine learning in {P}ython.
\newblock {\em Journal of Machine Learning Research}, 12:2825--2830, 2011.

\bibitem{libsvm}
Chih-Chung Chang and Chih-Jen Lin.
\newblock {LIBSVM}: A library for support vector machines.
\newblock {\em ACM Transactions on Intelligent Systems and Technology},
  2:27:1--27:27, 2011.
\newblock Software available at http://www.csie.ntu.edu.tw/~cjlin/libsvm

\bibitem{CPRDAT}
Wolfgang~J. Kern, Simon Orlob, and Martin Holler.
\newblock Cprdat, September 2021.
\newblock Zenodo-Repository.

\bibitem{gupta:2021}
Vishal Gupta, Robert~H Schmicker, Pamela Owens, Ava~E Pierce, and Ahamed~H
  Idris.
\newblock Software annotation of defibrillator files: ready for prime time?
\newblock {\em Resuscitation}, 160:7--13, 2021.

\bibitem{JAUREGUIBEITIA:2021}
Xabier Jaureguibeitia, Elisabete Aramendi, Unai Irusta, Erik Alonso, Tom~P.
  Aufderheide, Robert~H. Schmicker, Matthew Hansen, Robert Suchting, Jestin~N.
  Carlson, Ahamed~H. Idris, and Henry~E. Wang.
\newblock Methodology and framework for the analysis of cardiopulmonary
  resuscitation quality in large and heterogeneous cardiac arrest datasets.
\newblock {\em Resuscitation}, 168:44--51, 2021.

\end{thebibliography}

\newpage
\section*{Appendix}
Tab. 1. Utstein-style data sheet for data base
\newline

\centering
\begin{tabular*}{.71\textwidth}{|c @{\extracolsep{\fill}}|}
\hline 
\textbf{Absence of signs of circulation and/or considered for resuscitation        }  \\ 
$n=871$  \\ 
\hline 
\end{tabular*} 

\centering 
\begin{tikzpicture}
    \draw [-latex,line width=.5mm](4.5,0) -- (0.,-2);
    \draw [-latex,line width=.5mm](4.5,0) -- (9.,-2);
\end{tikzpicture}

\begin{minipage}{.48\textwidth}
\begin{tabular*}{\textwidth}{|p{.52\textwidth} p{.17\textwidth} p{.14\textwidth}|}
\hline 
\multicolumn{3}{|l|}{\textbf{Resuscitation not attempted}}\\
All cases & $n=2$&$ (0.2 \%)$\\
\hline
\end{tabular*}
\vspace{.5cm}

\begin{tabular*}{\textwidth}{|p{.52\textwidth} p{.17\textwidth} p{.14\textwidth}|}
\hline 
\multicolumn{3}{|l|}{\textbf{Location of Arrest}}\\
Home / Home for the aged & $n=619$&$ (71.1 \%)$\\
Public place & $n=161$ & $(18.5 \%)$\\
Other & $n=90 $& $(10.3 \%)$\\
Unknown & $n=1$ & $ (0.1 \%)$\\

\hline
\end{tabular*}
\vspace{.5cm}

\begin{tabular*}{\textwidth}{|p{.52\textwidth} p{.17\textwidth} p{.14\textwidth}|}
\hline 
\multicolumn{3}{|l|}{\textbf{Arrest witnessed/monitored}}\\
By layperson/bystander & $n=424 $&$ (48.7 \%)$\\
By healthcare personnel & $n=85 $&$ (9.8 \%)$ \\
Arrest not witnessed & $n=361 $&$ (41.4 \%)$\\
Unknown & $n=1$&$ (0.1 \%)$\\

\hline
\end{tabular*} 
\vspace{.5cm}

\begin{tabular*}{\textwidth}{|p{.52\textwidth} p{.17\textwidth} p{.14\textwidth}|}
\hline 
\textbf{CPR before EMS arrival} & $n=402 $&$ (46.2 \%)$\\
\hline
\end{tabular*} 
\vspace{.5cm}

\begin{tabular*}{\textwidth}{|p{.52\textwidth} p{.17\textwidth} p{.14\textwidth}|}
\hline 
\multicolumn{3}{|l|}{\textbf{Etiology}}\\
Presumed cardiac & $n=593 $&$ (68.1 \%)$\\
Trauma & $n=19 $&$ (2.2 \%)$\\
Submersion & $n=5 $&$ (0.6 \%)$\\
Respiratory & $n=124$ &$(14.2 \%)$\\
Other noncardiac & $n=73$&$(8.4 \%)$\\
Unknown & $n=57$&$(6.5 \%)$\\
\hline
\end{tabular*} 

\end{minipage}
\hfill
\begin{minipage}{.48\textwidth}
\begin{tabular*}{\textwidth}{|p{.52\textwidth} p{.17\textwidth} p{.14\textwidth}|}
\hline 
\multicolumn{3}{|l|}{\textbf{Resuscitation attempted}}\\
All cases & $n=869$&$ (99.8 \%)$\\
Any defibrillation & &\\
\qquad Yes &  $n=377$ &$ (43.3 \%)$ \\
\qquad No &  $n=274$ &$ (31.4 \%)$\\
\qquad Unknown &  $n=220$ &$ (25.2 \%)$ \\

Chest compressions & $n=869$ &$ (99.8 \%)$\\
Assisted ventilation & &\\
\qquad Yes &  $n=834$ &$ (95.8 \%)$\\
\qquad No &  $n=4$ &$ (0.5 \%)$\\
\qquad Unknown &  $n=33$ &$ (3.8 \%)$\\
\hline
\end{tabular*} 
\vspace{.5cm}

\begin{tabular*}{\textwidth}{|p{.52\textwidth} p{.17\textwidth} p{.14\textwidth}|}
\hline 
\multicolumn{3}{|l|}{\textbf{First monitored rhythm}}\\
Shockable & $n=258$&$ (29.6 \%)$\\
\qquad VF &  $n=258$ &$ (29.6 \%)$\\
\qquad VT &  $n=0$ &$ (0.0 \%)$\\
Non-Shockable & $n=611$ &$ (70.1 \%)$\\
\qquad Asystole &  $n=427$ &$ (49.0 \%)$\\
\qquad PEA &  $n=182$ &$ (20.9 \%)$\\
\qquad Unknown &  $n=2$ &$ (0.2 \%)$\\
\hline
\end{tabular*} 
\vspace{.5cm}

\begin{tabular*}{\textwidth}{|p{.52\textwidth} p{.17\textwidth} p{.14\textwidth}|}
\hline 
\multicolumn{3}{|l|}{\textbf{Outcome} (recorded for all categories)}\\
Any ROSC &   & \\
\qquad Yes &  $n=473$ &$ (54.3 \%)$\\
\qquad No &  $n=397$ &$ (45.6 \%)$\\
\qquad Unknown &  $n=1$ &$ (0.1 \%)$\\
\hline
\end{tabular*} 
\vfill
\end{minipage}

\begin{table*}
\centering
\caption{List of all features including their mean normalized importance from three different classifiers (Decision Tree, Random Forest and Lasso-regularized Logistic Regresssion). (StdDev \ldots standard deviation, PSD \ldots Power spectral density, ACC-ensemble-avg. \ldots ACC-ensemble-average }
\begin{tabular}{|c|l|c|l|c|}
\hline
\textbf{No}. & \textbf{Feature description} & \textbf{Signal Input} & \textbf{Reference} & \textbf{Importance} \\
\hline
$v_{15}$ & Peak-to-Peak amplitude of ACC-ensemble-avg.  & ACC & Ashouri et al, 2017 & 0.711  \\
$v_{8}$ & Windowed ACC Correlation & ACC \& ECG & this work & 0.629 \\
$v_{1}$ &  Root Mean Square value of ACC & ACC &  this work & 0.432\\
$v_{48}$ & \texttt{AMSA} & ECG & Elola et al, 2019 &  0.383 \\ 
$v_{11}$ &  Root Mean Square value of ACC-ensemble-avg. & ACC & Ashouri et al, 2017 & 0.339  \\
$v_{7}$ & Second derivative of ACC-correlation & ACC \& ECG & this work & 0.319  \\
$v_{36}$ & \texttt{meanRR} & ECG & Elola et al, 2020 &  0.273\\
$v_{49}$ & \texttt{HfP} & ECG & Elola et al, 2019 &  0.262\\
$v_{35}$ & $P_{h}$ & ECG & Ayala et al, 2014  &  0.208 \\
$v_{30}$ & $P_{LEA}$ & ECG & Ayala et al, 2014  & 0.183  \\
$v_{9}$ & Windowed-ECG Correlation & ECG & this work & 0.176  \\
$v_{45}$ & Mean of the absolute first difference of ECG & ECG & Elola et al, 2019 & 0.167  \\
$v_{44}$ & \texttt{StdSnorm} & ECG & Elola et al, 2020 &  0.167 \\
$v_{31}$ & $L_{min}$ & ECG & Ayala et al, 2014 & 0.146 \\
$v_{46}$ & StdDev of the absolute first difference of ECG & ECG & Elola et al, 2019 & 0.146  \\
$v_{32}$ & $b_S$ & ECG  & Ayala et al, 2014 & 0.138  \\
$v_{6}$ & ACC-correlation at ECG maximum & ACC \& ECG  & this work & 0.130  \\
$v_{4}$ & ECG-correlation maximum & ECG & this work &  0.124 \\
$v_{42}$ & \texttt{SlopeQRS} & ECG & Elola et al, 2020 &  0.114 \\
$v_{43}$ & \texttt{MSnorm} & ECG  & Elola et al, 2020 & 0.110 \\
$v_{17}$ & PSD of ACC in 0-3Hz band & ACC & Ashouri et al, 2017 & 0.100  \\
$v_{2}$ & Root Mean Square value of ECG& ECG & this work & 0.098 \\
$v_{38}$ & \texttt{MeanPP} & ECG & Elola et al, 2020  & 0.088  \\
$v_{10}$ & Quotient of windowed correlation & ACC \& ECG  & this work & 0.086  \\
$v_{3}$ & ACC-correlation maximum & ACC & this work &  0.063 \\
$v_{29}$ & Spectral Entropy of ACC & ACC  & Solar et al, 2017 & 0.059  \\
$v_{34}$ & $P_{fib}$ & ECG & Ayala et al, 2014  &  0.059 \\
$v_{40}$ & Mean QRS-width & ECG  & Elola et al, 2020 & 0.054  \\
$v_{13}$ & Skewness of ACC-ensemble-avg.   & ACC & Ashouri et al, 2017 & 0.054  \\
$v_{37}$ & \texttt{VarRR} & ECG & Elola et al, 2020 & 0.053  \\
$v_{16}$ & Peak-to-peak / power ratio of ACC-ensemble-avg.  & ACC & Ashouri et al, 2017 & 0.049  \\
$v_{33}$ & $n_P$ & ECG & Ayala et al, 2014 &  0.047 \\
$v_{39}$ & \texttt{StdPP} & ECG & Elola et al, 2020  & 0.041  \\
$v_{19}$ & PSD of ACC in 6-9Hz band & ACC & Ashouri et al, 2017 & 0.039  \\
$v_{47}$ & Kurtosis of ECG slope & ECG  & Elola et al, 2019  &  0.034 \\
$v_{21}$ & PSD of ACC in 12-15Hz band & ACC & Ashouri et al, 2017 & 0.033 \\
$v_{41}$ & StdDev QRS-width & ECG &  Elola et al, 2020  &  0.033 \\
$v_{5}$ & Spectral overlap & ACC \& ECG & this work &  0.032\\
$v_{12}$ & Kurtosis of ACC-ensemble-avg.  & ACC & Ashouri et al, 2017 & 0.030  \\
$v_{23}$ & Mean PSD of ACC & ACC & Ashouri et al, 2017 & 0.025 \\
$v_{20}$ & PSD of ACC in 9-12Hz band & ACC & Ashouri et al, 2017 & 0.021  \\
$v_{22}$ & PSD of ACC in 15-18Hz band & ACC & Ashouri et al, 2017 &  0.018 \\
$v_{18}$ & PSD of ACC in 3-6Hz band & ACC & Ashouri et al, 2017 & 0.016  \\
$v_{26}$ & Skewness of PSD of ACC & ACC & Ashouri et al, 2017 & 0.008 \\
$v_{24}$ & StdDev of PSD of ACC & ACC & Ashouri et al, 2017 & 0.007 \\
$v_{27}$ & Maximum of PSD of ACC & ACC & Ashouri et al, 2017 & 0.006 \\
$v_{25}$ & Kurtosis of PSD of ACC & ACC & Ashouri et al, 2017 & 0.005 \\
$v_{14}$ & Median of ACC-ensemble-avg.  & ACC & Ashouri et al, 2017 & 0.005  \\
$v_{28}$ & Frequency of maximum of PSD of ACC & ACC & Ashouri et al, 2017 & 0.003  \\
\hline
\end{tabular}
\end{table*}
\newpage
\newpage
\subsection*{Supplemental material: Feature importance of pure ECG-features}
\begin{figure}[h]
\centering
\includegraphics[trim=1.4cm 0.2cm 1.8cm 1.3cm,clip,width=.5 \textwidth]{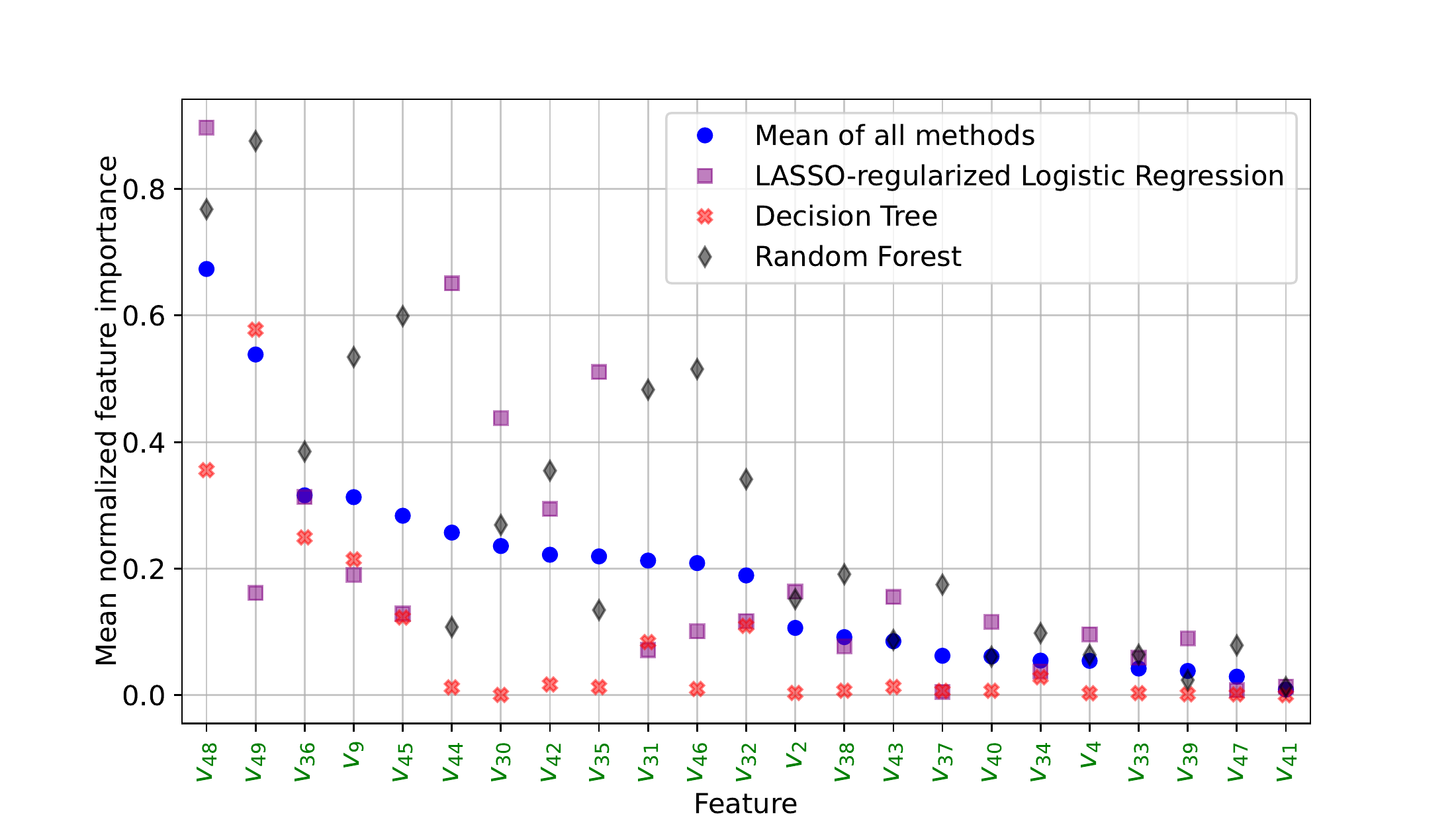}
\caption{Features importance of Lasso-regularized Logistic Regression, Decision Trees and Random Forests trained purely on ECG-features and their mean.}
\label{fig:feat_importance_ECG}
\end{figure}

We have also performed the feature importance analysis from subsection III.C using only ECG-features.
The results in Fig. \ref{fig:feat_importance_ECG} show that the most important features are similar as in the ACC+ECG-feature case, illustrating the importance of certain ECG features for classification signal.  Training an exemplary kernelized SVM with only the 10 best-ranked ECG-features yields a classifier with the follwing performance: (balanced accuracy: $0.769, (0.720,0.822)$, sensitivity: $0.838, (0.710,0.961)$, specificity: $0.700, (0.580,0.822)$, MCC: $0.544, (0.430,0.656)$, $F_1$: $0.820, (0.750,0.893)$). Compared to the classifier with the 10 best-ranked ACC- and ECG-features, the ECG-only classifier exhibits a slightly higher sensitivity, but a considerably lower specificity, resulting in an inferior overall performance.
\end{document}